\documentclass[traditabstract]{aa} 
\newcommand{\kms}{$\mathrm{km\, s^{-1}\, }$}
\newcommand{\msun}{M_{\odot}}

\usepackage{natbib}
\usepackage{graphicx}
\usepackage{longtable}
\usepackage{amsmath} 
\usepackage{color}

\usepackage{txfonts}
\begin{document}
   \title{A re-evaluation of the central velocity-dispersion profile in NGC~6388}
   
   \titlerunning{A re-evaluation of the central velocity-dispersion profile in NGC~6388}

   \author{Nora L\"utzgendorf
          \inst{1}
          \and
          Karl Gebhardt \inst{2}
          \and
          Holger Baumgardt \inst{3}
          \and 
          Eva Noyola \inst{2}
          \and 
          Nadine Neumayer \inst{4}   
          \and 
          Markus Kissler-Patig \inst{5} 
          \and 
          Tim de Zeeuw \inst{6,7}                 
          }

   \institute{ESA, Space Science Department,
              Keplerlaan 1, NL-2200 AG Noordwijk
              the Netherlands
         \and
			 Department of Astronomy, University of Texas at Austin, 
			 Austin, TX 78712, USA 
		\and
			 School of Mathematics and Physics,
			 University of Queensland,
			 Brisbane, QLD 4072, Australia
		\and
			 Max-Planck-Institute for Astronomy, 
			 K\"onigstuhl 17, 69117
			 Heidelberg, Germany	 
		\and
             Gemini Observatory, Northern Operations Center, 
             670 N. A'ohoku Place
			 Hilo, Hawaii, 96720, USA	
		\and
			 European Southern Observatory (ESO),
             Karl-Schwarzschild-Strasse 2, 85748 Garching, Germany	
        \and 
             Sterrewacht Leiden, 
             Leiden University, 
             Postbus 9513, 2300 RA, Leiden, 
             The Netherlands
}

   \date{Received xxxxxxx xx, xxxx; accepted xxxxx xx, xxxx}

 
  \abstract
   {The globular cluster NGC 6388 is among the most massive clusters in our Milky Way and has been the subject of many studies. Recently, two independent groups found very different results when measuring its central velocity-dispersion profile with different methods. While L\"utzgendorf et al. (2011) found a rising profile and a high central velocity dispersion ($23.3 $ \kms), measurements obtained by Lanzoni et al. (2013) showed a value 40\% lower. The value of the central velocity dispersion has a serious impact on the mass and possible presence of an intermediate-mass black hole at the center of NGC 6388. 
  The goal of this paper is to quantify the biases arising from measuring velocity dispersions from individual extracted stellar velocities versus the line broadening measurements of the integrated light using new tools to simulate realistic IFU observations.
   We use a photometric catalog of NGC 6388 to extract the positions and magnitudes from the brightest stars in the central three arcseconds of NGC 6388, and create a simulated SINFONI and ARGUS dataset. The construction of the IFU data cube is done with different observing conditions (i.e., Strehl ratios and seeing) reproducing the conditions reported for the original observations as closely as possible. In addition, we produce an N-body realization of a $\sim 10^6 M_{\odot}$ stellar cluster with the same photometric properties as NGC 6388 to account for unresolved stars.
   We find that the individual radial velocities, i.e. the measurements from the simulated SINFONI data, are systematically biased towards lower velocity dispersions. The reason is that due to the wings in the point spread function (PSF) of adaptive optics (AO) corrected data sets, the velocities get biased towards the mean cluster velocity. This study shows that even with AO supported observations, individual radial velocities in crowded fields do not reproduce the true velocity distribution. The ARGUS observations do not show this kind of bias but were found to have larger uncertainties than previously obtained. We find a bias towards higher velocity dispersions in the ARGUS pointing when fixing the extreme velocities of the three brightest stars but find those variations are within the determined uncertainties. We rerun Jeans models and fit them to the kinematic profile with the new uncertainties. This yields a black-hole mass of $M_{\bullet} = (2.8 \pm 0.4) \times 10^4 M_{\odot}$ and M/L ratio $M/L = (1.6 \pm 0.1) M_{\odot}/L_{\odot}$, consistent with our previous results.}

   \maketitle

   \maketitle
%

\section{Introduction}


Measuring central velocity-dispersion profiles in Galactic globular clusters aims to detect the crucial rise that would reveal the presence of an intermediate-mass black hole (IMBH) in the center. In the past years, several techniques have been established to derive the central velocity-dispersion profile in globular clusters. Starting with long-slit observations \citep[e.g.][]{peterson_1989} to measure individual velocities and integrated light, integral field units (IFUs) soon became a valuable tool to measure the central kinematics. \citet{dubath_1994} performed extensive numerical simulations of the velocity-dispersion determination from integrated light for the globular cluster M15. They found that integrated light measurements over small areas ($\sim 1 $ arsec$^2$) suffer from large statistical errors due to shot noise of a few massive stars, but that a larger coverage of the region of interest reproduces well the underlying velocity distribution. Besides using the integrated spectra of an IFU for the velocity dispersion determination, there is also the possibility of obtaining individual velocities of the stars in the field-of-view. This has been done using adaptive optics (AO) supported observations and extracting the spectra from the central spaxel of each star \citep[e.g.][]{lanzoni_2013} or by extracting individual spectra using a sophisticated deconvolution technique developed by \citet{kamann_2013}.

The first detection of an IMBH in a globular cluster using IFU spectroscopy was reported by \citet{noyola_2008} who studied the central region of the globular cluster $\omega$ Centauri and found the velocity dispersion to be rising towards the center. This detection, however, was challenged by \cite{vdMA_2010} who did not see such a rise in their velocity-dispersion profile when using proper motions and a different photometric center. After this, many IFU observations of globular clusters followed \citep{nora11, nora13, feldmeier_2013} which showed signatures of IMBH in the center in some Galactic globular clusters.  However, contradictory results using individual radial velocities instead of IFUs \citep{lanzoni_2013} or proper motions \citep{mcnamara_2012} kept the discussions going on whether or not IMBHs in the centres of globular clusters exist. Furthermore, X-ray and radio observations of the central regions of globular clusters result in no detection and low upper limits on the black-hole mass \citep[e.g.][]{maccarone_2005,strader_2012}. To date, there is no evidence of accreting IMBHs in globular clusters. The upper limits on black-hole masses from these observations are, however, very dependent on the knowledge of gas densities and accretion variability.

The case of NGC 6388 is an excellent example of contradicting measurements with different techniques. \citet{noyola_2006} found a shallow cusp in the central region of the surface brightness profile of NGC 6388. N-body simulations showed that this is expected for a cluster hosting an intermediate-mass black hole \citep{baumgardt_2005, noyola_2011}. However, \citet{vesperini_2010} argued that shallow cusps can also be produced by clusters that are in the process of core collapse and do not exhibit IMBHs, and the interpretation of the cusp is therefore still under debate. \cite{lanzoni_2007} investigated the projected density profile and the central surface brightness profile with a combination of HST high-resolution and ground-based wide-field observations. They found the observed profiles are well reproduced by a multimass, isotropic, spherical \citet{king_1966} model, with an added central black hole with a mass of $\sim 5.7 \times 10^3 M_{\odot}$. Our group presented the first central kinematic measurements of NGC 6388 in \citet{nora11} and found the velocity dispersion profile rising in the center up to $23.3$ \kms indicating the presence of an IMBH with a mass of $\sim 10^4 \msun$. However, \citet{lanzoni_2013} recently claimed the velocity dispersion in the center of NGC 6388 to be lower by 40\%. Using individual radial velocities extracted from the adaptive optics supported IFU of the SINFONI instrument, they found the velocity-dispersion to be 13.2 \kms using 52 individual velocities instead of 23.3 \kms from the ARGUS measurements. This high discrepancy is worrisome, as it affects the measurement on the existence and mass of a possible intermediate-mass black hole in the center.

Our group has therefore set out to investigate the effect on the velocity dispersion measurement when using individual velocities and integrated light. In Section \ref{sec:data} we introduce the two different observations and data sets. Section \ref{sec:simu} describes the simulations that were done to reconstruct the two different IFU observations and Section \ref{sec:extract} presents the extraction of the kinematics. The results of the Monte Carlo simulations and a summary can be found in Section \ref{sec:res} and \ref{sec:sum}, respectively. 
  
   \begin{figure}
  \centering
   \includegraphics[width=0.5\textwidth]{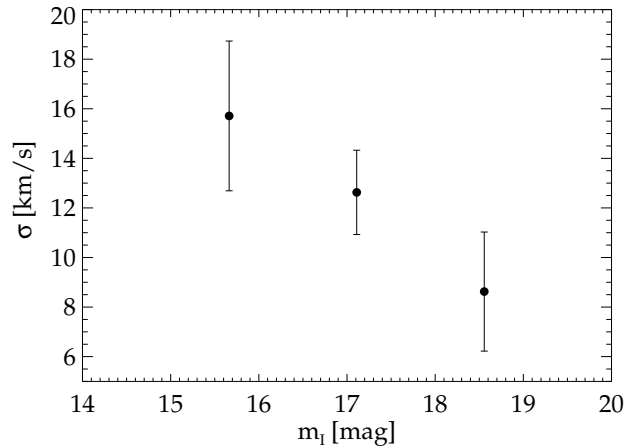}
      \caption{Velocity dispersion as a function of magnitude for the \citet{lanzoni_2013} sample. The sample shows a trend of decreasing velocity dispersion with increasing magnitude which cannot be explained by mass segregation.}
         \label{fig:dispmag}
   \end{figure}  
  

\section{The data sets}\label{sec:data}
 
Two independent data sets provide central velocity dispersions for NGC 6388. The first one was observed with the GIRAFFE spectrograph of the FLAMES (Fiber Large Array Multi Element Spectrograph) instrument at the Very Large Telescope (VLT) using the ARGUS mode \cite[Large Integral Field Unit,][]{pasquini_2002}. The second observation covered a smaller area in the center using the high spatial resolution capabilities of SINFONI \citep{eisenhauer_2003, bonnet_2004}, a near-IR (1.1-2.45 $\mu$m) integral field spectrograph fed by an adaptive optics module also mounted on the VLT. In this section we shortly describe the two observations and their conditions. 

The ARGUS observations were performed during two nights (2009-06-14/15, ESO proposal ID: 083.D-0444; PI: Noyola) with an average seeing of $0.8''$ (FWHM). The IFU unit was set to the 1 : 1 magnification scale (pixel size: $0.3 ''$, $14 \times 22$ pixel array) with the LR8 filter ($820-940 \, \mathrm{nm}, \, \mathrm{R} = 10400$) and pointed to three different positions to cover the central area of NGC 6388. The kinematics were obtained from the analysis of the Calcium Triplet ($\sim 850 \, \mathrm{nm}$) which is a strong absorption feature in the spectra. To compute the velocity dispersion profile we divided the pointing into six independent angular bins. In each bin, all spectra of all exposures where combined with a sigma clipping algorithm to remove any remaining cosmic rays. Velocity and velocity-dispersion profiles were computed using the penalized pixel-fitting (pPXF) program developed by \cite{cappellari_2004}.

The SINFONI observations were carried out between 2008 April and June (ESO proposal ID: 381.D-0329(A); PI: Lanzoni), under an average seeing of $\sim 0. 8''$ (FWHM). The instrument was set to the 100 mas plate scale (pixel size: $0.05 ''$, $3.2''\times 3.2''$ field-of-view) using the K-band grating ($1.95 - 2.45 \mu $m, $R = 4000$). The adaptive optics were performed using an R=12 magnitude star located $\sim 9''$ away from the cluster center. This results in a Strehl ratio (the amount of light contained in the diffraction-limited core of the PSF, with respect to the total flux) of $\sim 30\%$. Velocities were extracted individually for each star by matching the SINFONI pointing with high-resolution HST observations obtaining an accuracy of better than 0.2 spaxel. The spectra for the individual stars were extracted from the centroid positions of each star. The radial velocities were measured from the CO band-heads using a Fourier cross-correlation method \citep{tonry_1979} as implemented in the \verb|fxcor| IRAF task. The velocity dispersion for the central region was then determined using these 52 individual velocities and using the Maximum Likelihood method described in \citet{walker_2006}.

Figure 11 in \citet{lanzoni_2013} clearly shows the large discrepancy between the two results from the different data sets. In their study they find a central velocity dispersion of $\sigma_p = 13.2 \pm 1.3$ \kms, which is $40\%$ lower than the result of the ARGUS observations ($23.3 \pm 3.1$ \kms). Both measurements, however seem to agree with the outer kinematics obtained from \citet{lanzoni_2013} and \citet{lapenna_2014} using GIRAFFE/FLAMES data. 

   \begin{figure}
   \centering
   \includegraphics[width=0.5\textwidth]{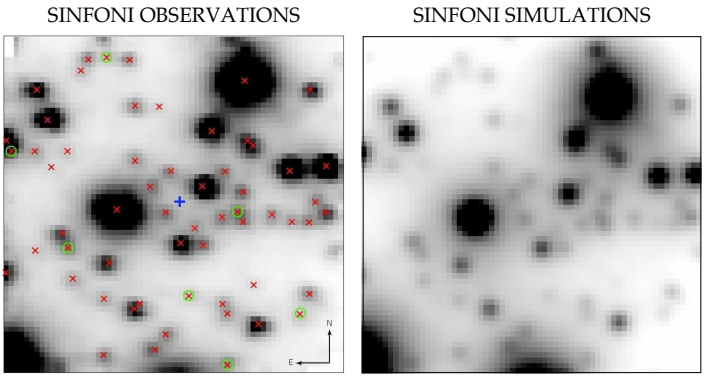}
      \caption{Observed (left) and simulated (right) $3.2''\times 3.2''$ field-of-view of the SINFONI IFU. The Strehl ratio of the simulated data cube is chosen to be the same as the observed one, 30\%. Red crosses in the left panel mark the stars that were detected in the SINFONI observations, while green circles mark the stars that were not used in the final analysis of \citet{lanzoni_2013} due to low signal-to-noise. }
         \label{fig:sinfoni}
   \end{figure}
   
      \begin{figure}
   \centering
   \includegraphics[width=0.5\textwidth]{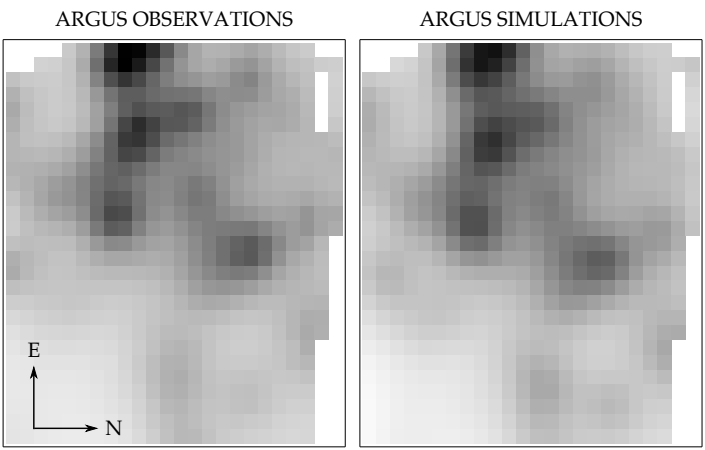}
      \caption{Observed (left) and simulated (right) $7.2''\times 8.1''$ field-of-view of the combined ARGUS IFU. The seeing of the simulated data cube is chosen to fit the observed data cube, $1.10''$.}
         \label{fig:sinfoni}
   \end{figure}
   
A concern is that the individual velocities as measured in \citet{lanzoni_2013} may be biased towards the cluster mean, resulting in a lower velocity dispersion. We address this by measuring the velocity dispersion in magnitude bins from their published data. There should be no difference in the velocity dispersion in these magnitude bins since the covered mass range is insignificant. If there is contamination from unresolved cluster light and light from the wings of nearby stars, we expect the effect to be larger for fainter stars resulting in the a lower dispersion value. Figure \ref{fig:dispmag} plots the velocity dispersion measured in three magnitude bins from their published data. It is clear that there is a trend to lower dispersion with magnitude signifying a potential bias. It is likely that even the brightest magnitude bin is also biased. By running a Spearman correlation test, using the IDL routine $r\_correlate$, on the absolute values of the individual velocities vs. magnitude we find a Spearman coefficient of $R= -0.29 \pm 0.04$ which indicates a slight anticorrelation with a two sided significance of $2.1 \sigma$. Thus, it appears that the \citet{lanzoni_2013} central velocity dispersion is biased low. This trend raises a concern but is no proof for a bias. For this reason we set out to test for a possible bias in both datasets in detail. Contamination from neighboring and unresolved stars is studied in detail in \cite{dubath_1994}. We explore a similar analysis to understand this bias, specific to NGC 6388.


\section{Simulated IFU observations}\label{sec:simu}

We created a tool to simulate IFU observations for arbitrary weather conditions, instrument setups and objects. A detailed description and verification of the code can be found in the upcoming paper L\"utzgendorf et al. (2015, in prep.). Here we describe the concept of the code and its specific application to NGC 6388 and the two observing techniques. 

Before creating the IFU dataset, a suitable star catalog is needed. To stay as close to NGC 6388 as possible we use the information on the brightest stars from a photometric catalog obtained in \citet{nora11} from an image taken with the Advanced Camera for Surveys (ACS) on the Hubble Space Telescope (HST). We use an N-body realization to add the influence of faint, unresolved stars to the catalog of NGC 6388. The model is set up by fitting an isochrone to the photometric catalog of NGC 6388. The best agreement is found with an age of 12 Gyr and a metallicity of $z=0.002$ (Bressan et al., 2012; Chen et al., 2014). The isochrone is used to create an evolved mass function and to obtain the stellar luminosity of individual stars. Using an isotropic \citet{king_1966} model, the distribution function with a central potential of $W_0 = 7$ (values taken from McLaughlin et al., 2012) is integrated to retrieve the density profile $\rho(r)$, the enclosed mass $M(<r)$, and the potential $\phi(r)$. The stars are distributed in space using $\rho(r)$ and a half-mass radius of $r_h = 1.5$ pc \citep{harris_1996}. As a final step, the distribution function and the potential are used to assign velocities to the individual stars in a way that the cluster fulfills virial equilibrium \citep[see][]{hilker_2007}. We adjust the number of stars in the simulated cluster such that we obtain the same number of bright stars in the central pointing as measured in the observations, to make sure that we do not overestimate the number of the fainter stars. The final cluster mass is $6.8 \times 10^5 M_{\odot}$, a bit lower than stated in the literature \citep[$\sim 10^6$,][]{mclaughlin_2005,nora11}. To add the observed stars to the catalog, we interpolate their positions to the isochrone to obtain temperatures, surface gravities and K-magnitudes. 
We also remove all stars from the catalog that are brighter than the faintest star from the SINFONI sample.
In each step we assign new velocities to the stars by drawing randomly from a velocity distribution centred around 0 \kms with a given velocity dispersion that we vary.

In order to construct a realistic data cube from the simulated data we use a set of synthetic spectra from the high-resolution synthetic stellar library \citep{Coelho_2005}  obtained from the VIZIER archive\footnote{Available at: http://vizier.cfa.harvard.edu/viz-bin/VizieR?-source=VI/120} for the ARGUS observations and synthetic spectra from \citet{husser_2013} for the infrared observations with SINFONI. These libraries include spectra that cover the Calcium Triplet region that is used for the ARGUS observations and the CO band heads to simulate SINFONI spectra. The spectra are first convolved with the spectral resolution of the respective instrument  (ARGUS: R = 10400, SINFONI: R=4000) and re-sampled to the wavelength range to match a number of spectral elements of a typical observed spectrum from each instrument.  With this grid of parameters each star in the cluster is assigned a spectrum with a spectral shift according to its radial velocity. As mentioned in \citet{lanzoni_2013}, the unresolved background is in general featureless because of the weakening of the CO band heads with lower temperatures. This effect is also visible in our data, since the stars get assigned a spectra according to their temperature and surface gravity. However, for completeness, we decided to keep the unresolved background even if it will not have a large effect on the measurements.


The IFU construction is done by a C program that efficiently parallelizes the tasks of taking each star, finding a spectral match for it, taking its position and flux to construct a Moffat point spread function (PSF) 

\begin{equation}
f(x,y; \alpha,\beta)=\left(\beta-1\right)\left(\pi\alpha^2\right)^{-1}\left[1+\left(\frac{x^2+y^2}{\alpha^2}\right)\right]^{-\beta},
\end{equation}

with $\beta = 2.5$ and $\alpha = FWHM / (2 \sqrt{2^{1/\beta} -1})$, on the IFU grid and combines the spectra of all stars in each spaxel weighted by their luminosity. 

In the case of AO observations we use the combination of two Moffat functions, one with an intrinsic FWHM of $0.15''$ and one with the native seeing (i.e. the observed seeing, 0.8$''$) weighted by the measured Strehl ratio. The construction of the final PSF for AO observations therefore follows the equation:

\begin{equation}
f(x,y; \alpha_i,\alpha_o,\beta)=S * f(x,y; \alpha_i,\beta) +  (1-S) * f(x,y; \alpha_o,\beta),
\end{equation}

where $S$ is the Strehl ratio and $\alpha_i$ and $\alpha_o$ are derived from the inner (intrinsic) and outer (native) FWHM, respectively. Furthermore, the routine applies noise to each spaxel. We request a minimum S/N of 10 for the faintest spaxel and apply noise to the remaining spaxels accordingly. Figure \ref{fig:sinfoni} compares the central pointing of the SINFONI observations with the simulated data cube. The visual similarity of the two data sets gives confidence in the performance of the IFU simulation routine.  The value for the intrinsic FWHM is in good agreement with the values found in \citet{rusli_2013}. In addition, we select the 11 brightest isolated stars from the SINFONI cube  and analyse their PSF. For each star we extract a 1D PSF profile and fit a two-component Moffat function. The free parameters are the FWHM of the inner component (FWHM$_i$), the FWHM of the outer component (FWHM$_o$), the Strehl ratio (i.e. the ratio of these two components), and the scale. The fits show that the intrinsic FWHM does vary between $0.13''$ and $0.15''$. The Strehl ratios range from 0.1 to 0.5, with the very brightest stars having the largest values. In general this shows that the parameters that we are using are a good representation of the observed data. Furthermore, we directly compare the 1D profiles of our simulated SINFONI data cube and the observed one and find very good agreement with the chosen values. 

The simulated PSFs assume two simplifications: Spherical symmetry and no variability across the field of view. The ellipticity that is observed in the measured datacube is difficult to reproduce in the simulations and would introduce even more free parameters. We concluded, however, that the effect of the elliptic shape of the PSF will be small to the overall analysis. The ellipticity will cause some of the stars (that are aligned with the minor axis of the ellipse) to be less contaminated by bright stars while other stars (aligned with the major axis) will be more contaminated. This would have an effect if we would make a star-by-star analysis. However, since we are computing the velocity dispersion of the system, these effects will cancel each other out. The same argument holds for possible variability of the PSF across the field of view.  The 1D fits of the PSF show that our models are compatible with the observations and that the fitted PSF parameters scatter around the parameters that we use for the final simulations.

    \begin{figure*}
   \centering
   \includegraphics[width=\textwidth]{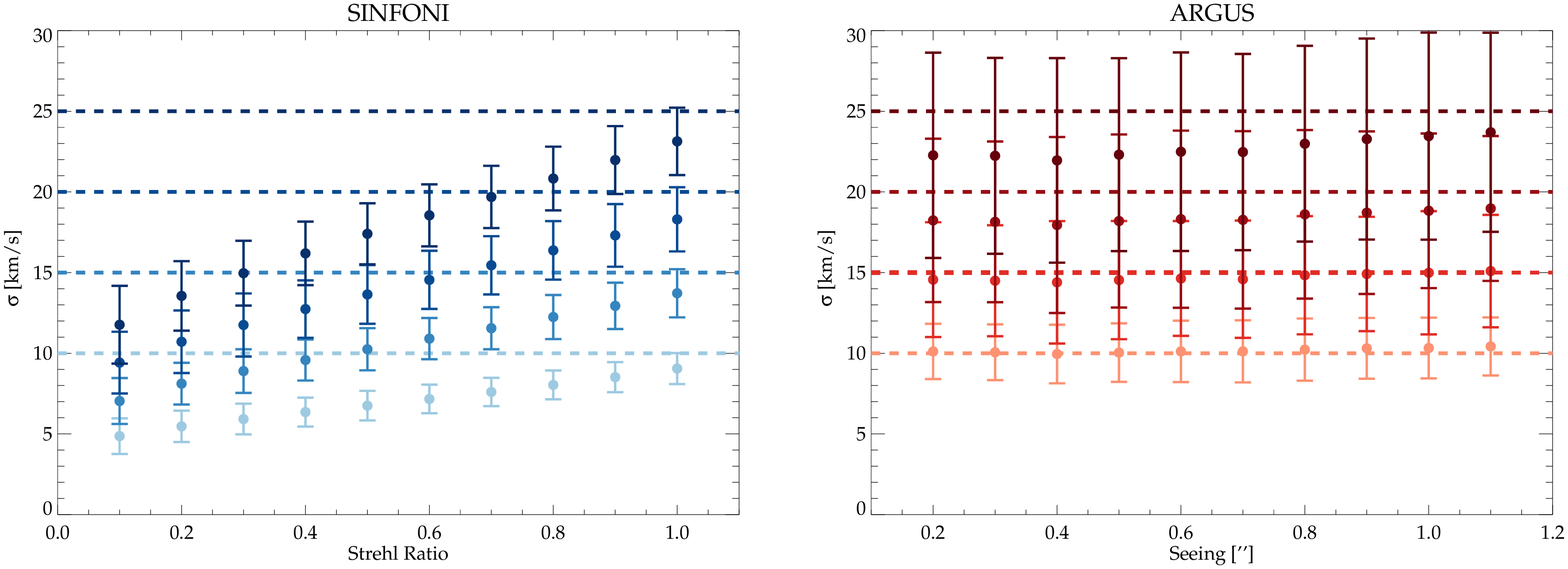}
      \caption{IFU simulations for SINFONI and ARGUS (central bin) with different input velocity dispersions as a function of Strehl ratio and seeing. The input velocity dispersions are shown as dashed lines, the measured velocity dispersions as dots in the corresponding color.}
         \label{fig:res}
   \end{figure*}

       \begin{figure*}
   \centering
   \includegraphics[width=\textwidth]{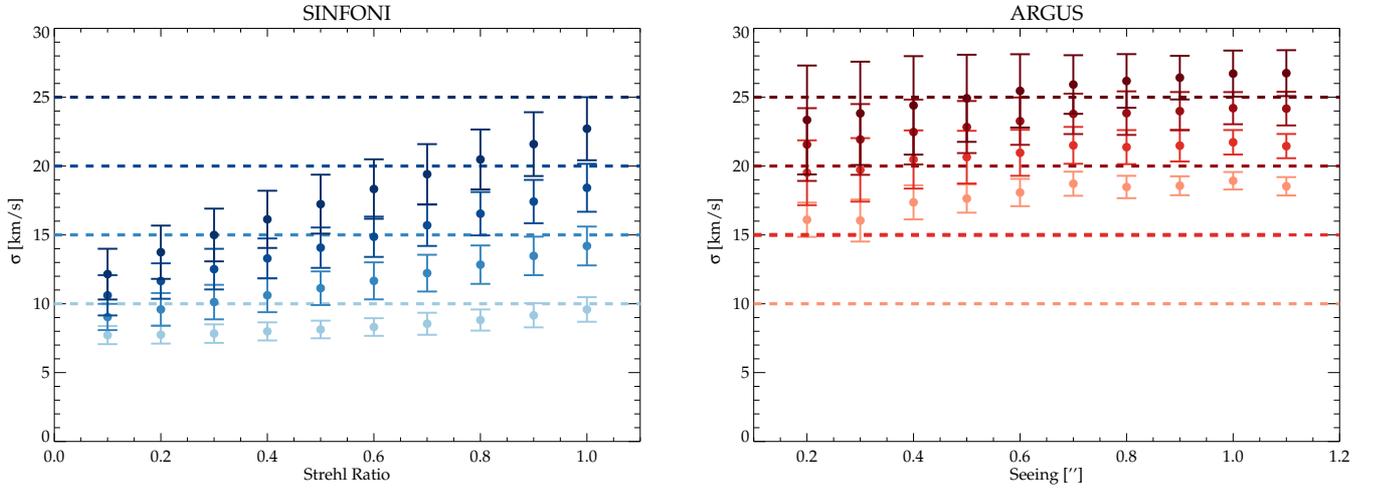}
      \caption{Same as Figure \ref{fig:res} while fixing the velocities of the three brightest stars to their measured velocities and different input velocity dispersions as a function of Strehl ratio and seeing.}
         \label{fig:res3stars}
   \end{figure*}

\section{Extraction of kinematics}\label{sec:extract}

The extraction of the kinematics is performed in a similar way as with the observed data. For SINFONI we use the star position from the catalog to extract the spectrum of the central spaxel of each star from the IFU. We note that by skipping the process of cross correlating with an independent catalog we neglect the errors that might arise from mismatches with the catalogs but we consider them as marginal since the IFU data cube is constructed using the same catalog.From the extracted spectra we then measure the velocities of each star using \verb|fxcor| from the IRAF package and a list of templates with the same stellar parameters as the stars. We make the assumption that the internal parameters of the stars are perfectly known and that we are using the perfect template for each star.
The final velocity dispersion is then computed by using  the Maximum Likelihood method introduced by \cite{pryor_1993} to account for the individual uncertainties on the velocity measurements. We repeat this measurement for IFU simulations with different Strehl ratios ranging from 10\% to 100\%. The results are discussed in the following section. 

For the simulated ARGUS observations we adopt the position of the center and the radial bins from \citet{nora11}. We then combine all spectra in each bin by applying an iterative sigma clipping scheme. The velocity dispersion is derived by performing fits with pPXF on the combined spectrum but this time not measuring the centroid (i.e. the velocity) but directly the velocity dispersion from the broadening of the lines. For both SINFONI and ARGUS simulations we repeat the procedure described above for 100 iterations in which we assign new velocities to all the stars in each step. This is performed for four different input velocity dispersions (10, 15, 20, 25 \kms) and 10 different values for seeing and Strehl ratios. The final values and their uncertainties are taken from the mean and the standard deviation of all iterations.

\section{Results}\label{sec:res}

\subsection{Velocity measurements}

Figure \ref{fig:res} shows the result of the two IFU simulations. For the SINFONI observations we show the obtained velocity dispersion as a function of Strehl ratio and for different input velocity dispersions. For the ARGUS observations the same is shown but for different values of seeing. The Figure shows that despite large uncertainties, the ARGUS measurements closely resemble the input velocity dispersions (dashed lines of the same color). The SINFONI observations, however, are strongly biased towards lower velocity dispersions. This can be explained by individual and especially faint stars being contaminated by the wings of their neighbouring stars. This pushes measured velocities closer to the mean velocity of the cluster and therefore lowers the velocity dispersion. As shown in Figure \ref{fig:res} this effect is especially severe for low Strehl ratios but even with a perfect Strehl ratio of 100\% the bias persists. We find that a measured velocity dispersion of about 13 \kms (as measured in \citet{lanzoni_2013} with a Strehl ratio of 30\%) requires an input velocity dispersion of $\sim$ 25 \kms. This is in excellent agreement with the measurements obtained in \citet{nora11} and would explain the discrepancy between the two measurements.

\subsection{Discussion of the bias}

\begin{figure}
   \centering
   \includegraphics[width=0.5\textwidth]{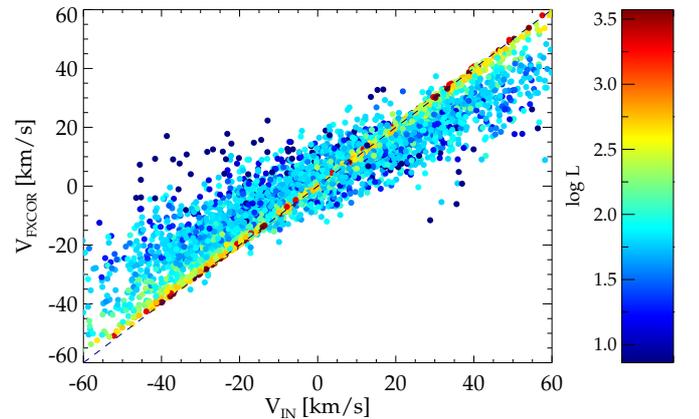}
      \caption{Measured velocities versus input velocities for 100 Monte Carlo runs on the SINFONI dataset. The stars are color coded according to their magnitudes.}
         \label{fig:vel}
   \end{figure}

The bias towards lower velocity dispersion in the SINFONI dataset is concerning. To investigate further, we plot the measured velocities versus the input velocities of the SINFONI stars for a set of 10 Monte Carlo runs. We also color code the stars by their magnitudes. Figure \ref{fig:vel} shows the measured velocities. It is obvious that the bias depends on the magnitude of each star and that for the very bright stars (dark red - red) the velocities are measured well, while for the faint stars (blue) the bias is always directed to a lower velocity. 

         \begin{figure*}
   \centering
   \includegraphics[width=\textwidth]{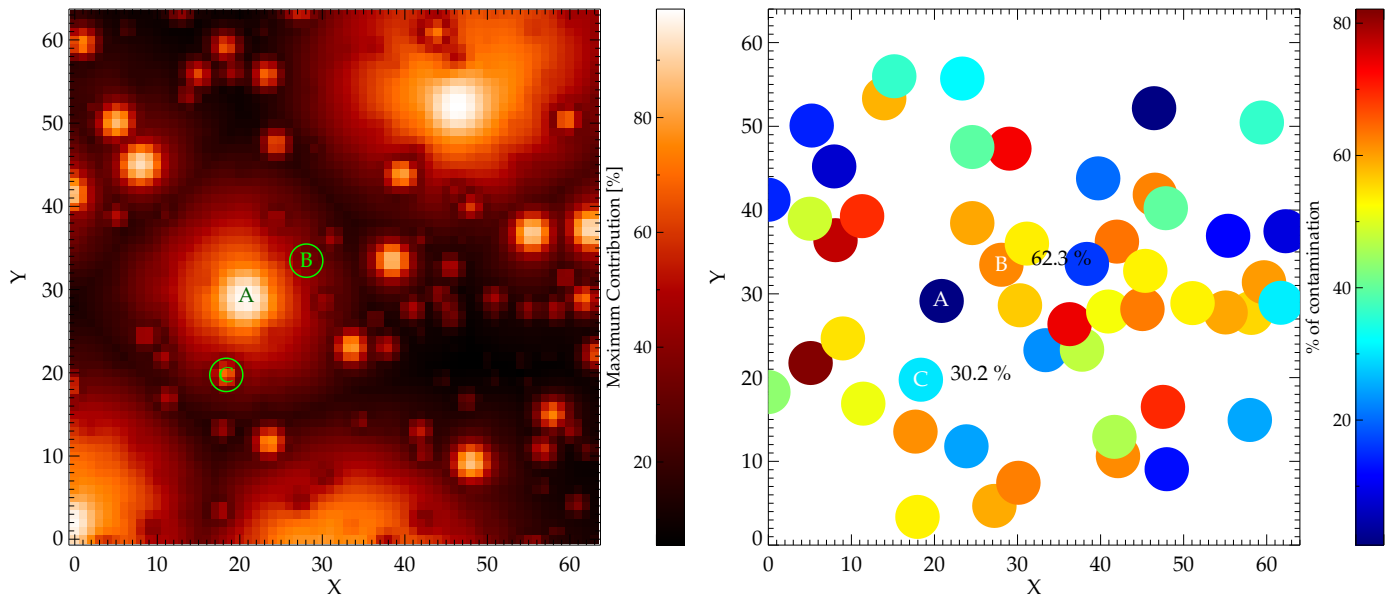}
      \caption{Contamination map for the SINFONI pointing. Left panel shows the maximum contribution of a single star to each spaxel. From this map the right panel is constructed that shows for each star that is measured in the SINFONI pointing the percentage of contamination by other stars. All stars above the dark blue regime are contaminated by more than 20\%.}
         \label{fig:cont}
   \end{figure*}

The reason for this bias can not be explained by the unresolved background, since, as discussed in the previous sections, this is featureless. However, the influence of the neighboring stars and the contamination of the faint stars by the wings of the bright stars must not be underestimated. To quantify this effect we compute a contamination map for each spaxel in the SINFONI pointing. Figure \ref{fig:cont} shows the maximum contribution to each spaxel by any star in the left panel. From this map we created the contamination map in the right panel where we examined the spaxel of the measured stars in the SINFONI pointing and determined to which level these spaxel (that were used to derive the velocity) are contaminated. To give an example we selected two stars close to the brightest star (A) with different magnitudes and state their contamination. While star B is already moderately contaminated (30\%) by the bright star and other stars around it, star C is highly contaminated by the wing of the very bright star A by a value of $\sim 60 \%$, i.e. it contributes less than half of the light to this spaxel. This shows the importance of the PSF wings even in AO observations and the contaminating effect of the bright stars. For a Gaussian velocity distribution, the contamination from the neighboring stars will bias the velocities of the faint stars to the mean velocity. We note that we also rerun the simulations using a smaller value for the outer FWHM (FWHM$_o = 0.5''$) in order to test the effect of the native PSF. We find the bias reduced by $10 - 15\%$ but still present and argue that a seeing of $0.5''$ is not representative of average VLT observations and much lower than the observed seeing of $0.8''$.

Compared to the SINFONI data set, the simulated ARGUS data in the right panel of Figure \ref{fig:res} does not show such a bias. However, the simulations of the innermost bin result in a very large scatter and an uncertainty of more than 20 \%. Similar results have been already reported by Bianchini et al. (2015, submitted).  The highest values for $\sigma$ also show a slight trend towards lower values, but still consistent with the input velocity dispersion, when taking the large uncertainties into account. This lower values most probably arise from the fact that the innermost bin is dominated by one or two single stars with narrow absorption lines.  Not shown in Figure \ref{fig:res} are the results for the five other bins of the ARGUS IFU that were also included in the simulations. The resulting uncertainties are smaller than for the central bin and are shown in Figure \ref{fig:comp1}.

In order to follow the argument by \citet{lanzoni_2013} that the brightest stars might bias the velocity dispersion towards higher values for this specific case of velocity distributions we run the same simulations as before but this time with fixing the velocities of the three brightest stars to the measured velocities of $v_1 = -23.2$ \kms, $v_2 = 18.4$ \kms, (measured by \citet{lanzoni_2013}, the two brightest stars in the sample) and $v_3 = -46.6$ \kms (measured from the ARGUS pointing and not present in the \citet{lanzoni_2013} dataset. The velocities of the rest of the stars are again drawn from a velocity distribution with different input velocity dispersion. Figure \ref{fig:res3stars} shows the result of these runs. While the SINFONI dataset is only slightly affected (compared to Figure \ref{fig:res}), the impact of the three bright stars having opposite velocities is visible in the ARGUS simulation. For low velocity dispersions the three stars drive the velocity dispersion up to 20 \kms but for high velocity dispersions there is barely a difference (while the bias in the SINFONI data is visible for all velocity dispersions). The reason for this directed bias lies in the fact that adding two very fast stars on top of a velocity distribution with a low velocity dispersion for all simulation runs is not a correct representation of this distribution. Further, we want to stress that a velocity dispersion as low as 18 \kms is still allowed by the uncertainties. As a last point it is worth mentioning that only the central point is heavily affected by this specific constellation of stars but not the bins further out. As we show in the next section, only the bins further out have an effect on the black hole mass in the Jeans modelling, due to the large errorbars on the central point. 

Thus, we conclude that even excluding the central region where the bright stars are, there still is a rise in the dispersion at larger radii where these bright stars have no influence. The observed rise in the dispersion is therefore not dependent on the bright stars. Second, one cannot remove the stars from the sample, unless there are reasons due to them not being in dynamical equilibrium. Our simulations are designed to measure the uncertainties on the dispersion, including the brightness of the stars. We must include them when measuring the dispersion, and then we determine the uncertainty from the simulations. It is irrelevant what their velocity distribution is for the simulations. Furthermore, we believe that the possible rotation signature in the center of NGC 6388 is not due to three bright stars but to a group of stars as already stated in \citet{nora11}. The reason why this is not seen in the dataset of \citet{lanzoni_2013}  is assumed to be due to the bias of the fainter stars towards the mean velocity which leads to a weakening of the rotation amplitude. While any possible rotation is interesting for the cluster kinematics, it must be included for a dynamical analysis for the enclosed mass. The second moment we measure from the integrated light naturally includes both the ordered and hot components of the stellar kinematics.

\subsection{New dynamical models}

      \begin{figure}
   \centering
   \includegraphics[width=0.5\textwidth]{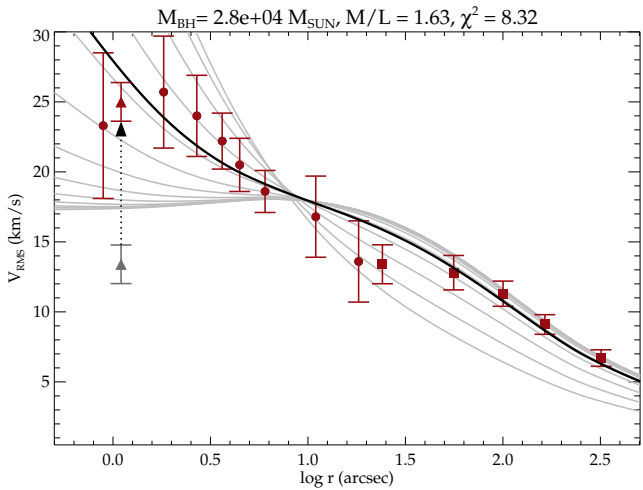}
      \caption{Jeans models for the velocity dispersion profile of NGC 6388 using the profile obtained in \citet{nora11} (dots), the uncertainties on these measurements obtained in this work, the corrected value from \citet{lanzoni_2013} (triangle), and the outer points obtained by \citet{lanzoni_2013} and \citet[][squares]{lapenna_2014} .}
         \label{fig:comp1}
   \end{figure}
   
          \begin{figure*}
   \centering
   \includegraphics[width=\textwidth]{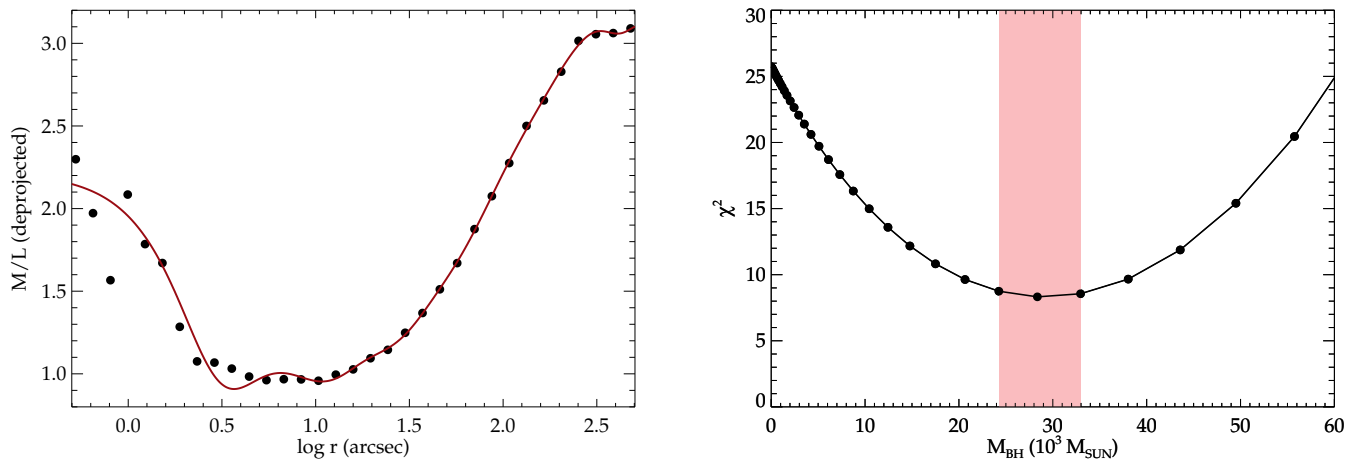}
      \caption{Input M/L profile obtained from N-body simulations (left) and $\chi^2$ for the different Jeans models from Figure \ref{fig:comp1} (right) with a best-fit black-hole mass of $M_{\bullet} = (2.4 \pm 0.4) \times 10^4 M_{\odot}$. While the rise in the M/L in the central region is due to stellar remnants, the rise at larger radii originates from low-mass stars.}
         \label{fig:comp2}
   \end{figure*}

As mentioned in the previous paragraph, we simulate the remaining five bins of the ARGUS observations in the same way as described above and find that the uncertainties in \citet{nora11} are underestimated (see Figure \ref{fig:comp1}). The difference might arise from the more advanced IFU modelling compared to the method used in \citet{nora11} where, for example, only one spectral type of star was used. We therefore rerun our Jeans models as described in \citet{nora11}, using the same, but slightly smoothed surface brightness profile in addition with a M/L profile obtained from N-body simulations shown in Figure \ref{fig:comp2}. The M/L profile is obtained by fitting a grid of N-body models to the photometric and kinematic data of NGC 6388 and computing the M/L for the best-fit model. The models are computed with 10\% stellar-mass black holes and neutron stars retention fraction and an initial \cite{kroupa_2001} mass function.  More details about the N-body models are presented in \citet{mcnamara_2012}.

For completeness we also add the most recently published outer datatpoints from \citet[][squares]{lanzoni_2013,lapenna_2014} and the corrected datapoint from \citet{lanzoni_2013} (triangle, correction is indicated with the dashed arrow, not included in the fit). Figure \ref{fig:comp1} shows Jeans models with different black-holes masses, the projected M/L profile and its MGE parametrization, as well as the $\chi^2$ values for each model are shown in Figure \ref{fig:comp2}. We note that our Jeans models differ from the ones of \citet{lanzoni_2013} by the M/L profile and result in a higher model central velocity dispersion even without a black hole. Furthermore, models used by \citet{lanzoni_2013} are the results of parametric multi-mass \citet{king_1966} and \citet{wilson_1975} models while our models only take the measured and smoothed surface brightness profile and a M/L profile as input.Furhtermore, it should be stated that the innermost datapoint does not influence the black-hole measurement due to its large uncertainty. However, the bins further out still show the clear rise and cannot be the result of contamination by the innermost bright stars. The models result in a black-hole mass of $M_{\bullet} = (2.8 \pm 0.4) \times 10^4 M_{\odot}$ and a total $M/L = (1.6 \pm 0.1) M_{\odot}/L_{\odot}$. We note that the uncertainties on the black-hole mass might be underestimated in this case since we do not perform Monte Carlo simulations on the surface brightness profile as in \citet{nora11}. Both the black-hole mass and the M/L agree well within their uncertainties with the results from \citet{nora11} ($M_{\bullet} = (1.7 \pm 0.9) \times 10^4 M_{\odot}, M/L = (1.6 \pm 0.3) M_{\odot}/L_{\odot}$).

We note that we are using simplified models to obtain the black-hole mass. By using isotropic Jeans models we are neglecting the fact that a globular cluster most likely shows anisotropy in its outer regions. In \citet{nora11} we demonstrated that in the central regions of a globular cluster anisotropy quickly disappears due to relaxation processes. By fitting the entire cluster profile, however, we include regions that can be affected by anisotropy \citep{zocchi_2015}. This can affect the overall shape of the model velocity dispersion. More sophisticated models such as $f_{\nu}$ \citep[e.g.][]{zocchi_2012,zocchi_2015} or Schwarzschild Models \citep[e.g.][]{bosch_2006, jardel_2013} would therefore be the better representation of such a complex system, but are outside of the scope of this paper.

\section{Summary}\label{sec:sum}

We perform IFU simulations of observations with two different IFUs for the particular case of the globular cluster NGC 6388. Using a combination of synthetic spectra, N-body realizations and observational data we reproduce observations of the central region of NGC 6388 as close to reality as possible. The kinematics are extracted in similar ways as done for the observed data. That allows a direct comparison of the biases coming with the technique of measuring individual velocities and line broadening of integrated light. The results are evaluated for different observing conditions and input velocity dispersions.

We show that both measurements feature large uncertainties but only the SINFONI datasets are clearly biased towards lower velocity dispersions depending on the Strehl ratio of the observations. The bias towards lower velocities can be explained by the contamination of neighbouring stars which shifts the velocities of the individual stars closer to the mean velocity. For a measured velocity dispersion of about 13 \kms, an input velocity dispersion of about 25 \kms is required. This is in good agreement with our measurements presented in \citet{nora11}. In the integrated light method on the other hand, stars and background light are both valuable signal that is used to derive the velocity dispersion. However, we noticed a bias towards higher velocity dispersions using integrated light when fixing the extreme velocities of the three brightest stars regardless of the input velocity dispersion. The bias is included in the large uncertainties of the innermost datapoint and disappears for high velocity dispersions. We run Jeans models on the whole kinematic dataset including the outer points derived in \citet{lanzoni_2013} and \citet{lapenna_2014}, and including the newly derived uncertainties for our points measured with the ARGUS IFU. The fits result in a black-hole mass of $M_{\bullet} = (2.8 \pm 0.4) \times 10^4 M_{\odot}$ and M/L ratio $M/L = (1.6 \pm 0.1) M_{\odot}/L_{\odot}$ which is in good agreement with the previous results on this data set. 

This analysis is specific for NGC 6388 where there are many observations and photometric data to compare with. However, biases for measurements with individual velocities are most likely present in all of the data sets that have been obtained so far. NGC 6388 belongs to the densest globular clusters in our Milky Way. Therefore it is understandable that the blend effects are severe in this environment. However, even with sparse clusters one has to be careful when taking individual velocities and tests for their reliability are always needed. For core-collapsed clusters with densities even higher than NGC 3688, it is not recommended to use this technique at all. One solution to the issue might arise from a new deconvolution technique developed by \citet{kamann_2013}. The precise extraction method reduces the overblend effects of neighboring and background stars to a minimum and allows to analyze data with much lower spectral resolution than the integrated method. 

For the case of NGC 6388 and other nearby clusters there are other methods for measuring the velocity dispersion. Proper motions for a large sample of Galactic globular clusters have been obtained in the frame of the ACS survey of Galactic globular clusters \citep{sarajedini_2007}. Using proper motions together with discrete Jeans modelling \citep[e.g.][]{watkins_2013} brings a vast advantage to globular cluster modelling. By including M/L profiles and anisotropy, these models do not just apply the physics needed to the models but also do not need to make assumptions for binning. \citet{watkins_2015} published the kinematic profiles of a large set of globular clusters including NGC 6388. Unfortunately their profile does not extend further inside than 1 arcsec and therefore does not cover the critical region that is discussed in this paper. If there will be enough usable data points on the central parts of NGC 6388 measured in the future, proper motions might help to solve the discrepancies in radial velocity measurements.

A detailed and more general study on the influence of IFU measurements for semi-resolved objects such as globular clusters as a function of cluster properties and observing conditions is crucial to produce reliable results. In terms of non-AO supported observations similar to the ARGUS data set and the deconvolution extraction technique by \citet{kamann_2013} we have performed detailed studies on the reliability of velocity dispersion measurements with IFUs and their effect on the determination of the total mass and the black hole mass in the center, if present. The results of this study will soon be published in L\"utzgendorf et al. (2015, in prep.). This provides a valuable tool for future and present observations and a key to calibrate measurements performed with IFU instruments.

\bibliographystyle{aa}
\bibliography{literatur}

\end{document}